\documentclass[twoside, conference]{IEEEtran}

\usepackage[normalem]{ulem}
\usepackage{tikz}
\usepackage{url}
\usepackage{pgfplots}
\usepackage{amsmath}
\usepackage{multirow}
\usepackage{threeparttable}
\usepackage{enumitem}
\usepackage{listings}
\usepackage{subfig}
\usepackage{amsthm}
\usepackage{braket}
\usepackage[ruled,linesnumbered,lined,boxed,commentsnumbered]{algorithm2e}
\usepgfplotslibrary{statistics}

\definecolor{mygreen}{rgb}{0,0.6,0}
\newcounter{theorem_counter}
\begin{document}

\title{\bf{qTask: Task-parallel Quantum Circuit Simulation with Incrementality}}
\author{Tsung-Wei Huang \\ 
tsung-wei.huang@utah.edu \\
Department of Electrical and Computer Engineering, University of Utah}

\maketitle

\begin{abstract}

Incremental quantum circuit simulation has emerged
as an important tool for simulation-driven quantum applications,
such as circuit synthesis, verification, and analysis.
When a small portion of the circuit is modified,
the simulator must incrementally update state amplitudes
for reasonable turnaround time and productivity.
However, this type of incrementality has been largely ignored 
by existing research.
To fill this gap,
we introduce a new incremental quantum circuit simulator called qTask.
qTask leverages a task-parallel decomposition strategy 
to explore both inter- and intra-gate operation parallelisms
from partitioned data blocks.
Our partitioning strategy effectively narrows down incremental update
to a small set of partitions affected by circuit modifiers.
We have demonstrated the promising performance of qTask on QASMBench benchmarks.
Compared to two state-of-the-art simulators, Qulacs and Qiskit,
qTask is respectively 1.46$\times$ and 1.71$\times$ faster for full simulation 
and 5.77$\times$ and 9.76$\times$ faster for incremental simulation.


\end{abstract}

\section{Introduction}

Quantum computing (QC) is a promising computing paradigm
for tackling certain types of problems that are classically intractable,
such as cryptography, chemistry simulation, and finance~\cite{QuantumBible}.
Among various QC applications,
classical \textit{quantum circuit simulation} (QCS) 
is essential for researchers to 
understand quantum operations, design quantum algorithms, 
and validate quantum circuit functionality~\cite{Wu_19_01}.
However, QCS is extremely challenging because it demands
large computation and memory to evaluate state amplitudes
of qubits.
For example, a full simulation of an $n$-qubit circuit 
requires an exponential size of vector to store $2^n$ amplitudes,
as a result of superposition.
To tackle this challenge,
QCS researchers have explored parallel computing~\cite{Qulacs, FasterQCS}, 
data compression~\cite{Wu_19_01}, 
circuit optimization~\cite{PatternMatchingOpt, QCCircuitOpt}, etc. 

\begin{figure}[!h]
  \centering
 \centerline{\includegraphics[width=1.\columnwidth]{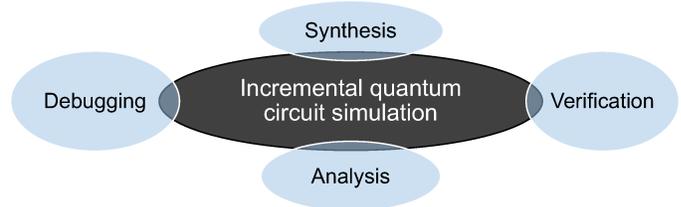}}
  \caption{
    Incremental quantum circuit simulation is a key enabler
    to high-performance simulation-driven quantum applications.
   }
  \label{fig::iqs_applications}
\end{figure}

Despite the rapid growth of QCS research,
existing simulators are largely short of a key feature--\textit{incrementality}.
Incremental QCS has recently 
emerged as an important tool for simulation-driven QC applications,
as shown in Figure \ref{fig::iqs_applications}.
For example,
quantum circuit synthesizers can iteratively modify 
circuit gates to increase certain state probability
and verify the results with thousands of simulation runs~\cite{DA4QC, Classiq, Gokhale_19_01, Gokul_23_01};
developers can issue step-by-step simulation calls
to debug how qubits change during the implementation 
of quantum algorithms;
equivalence checking tools can repetitively add or remove gates 
to verify how similar two circuits are based on simulation results~\cite{QuantumEquivalenceChecking}.
For these applications,
%
when small portions of a quantum circuit is modified,
re-simulating the full circuit is infeasible 
from a turnaround time and productivity perspective.
The simulator must incrementally update only affected regions
and ensure state integrity in an efficient manner.

There are several challenges for designing
an efficient incremental QCS system.
First,
running QCS in a static environment is very different from
a dynamic environment. 
When a quantum circuit begins to change, it can become very difficult to
reorganize data structures and keep algorithmic invariants consistent 
over incremental operations.
Second,
achieving fast incremental QCS requires very strategic task partitioning
to make the most of parallelism~\cite{QuantumCircuitPartitioning}.
When applications modify the circuits,
we need to quickly identify affected partitions and restructure their task dependencies
for incremental update.
%
Last but not least, 
although algorithms of incrementality have been widely studied in the classical design flow
of digital circuits (e.g., incremental timing/power analysis~\cite{Huang_16_01, OpenTimerv2}),
we cannot directly reuse them due to distinct behavior of quantum circuits,
such as superposition and entanglement.

To overcome these challenges,
we introduce \textit{qTask}, a state vector-based quantum circuit simulator that efficiently 
supports incrementality using task parallelism.
To the best knowledge of authors,
qTask is the first incremental quantum circuit simulator
in the literature.
Our result can largely benefit many simulation-driven quantum applications.
We summarize our technical contributions as follows:

\begin{itemize}[leftmargin=*]\itemsep=2pt

\item We present a lightweight C++ programming model to 
support incremental QCS.
Applications can use our circuit modifiers
to modify existing quantum circuits and call state update to 
transparently perform incremental simulation.

\item We present a task graph-based partitioning strategy to explore
both inter- and intra-gate operation parallelisms from a quantum circuit.
Our strategy parallelizes both full simulation 
and incremental simulation.

\item We present an efficient technique to maintain invariants 
of our task partitioning over sequences of circuit modifiers.
When a state update call is issued,
we can quickly identify affected partitions and restructure
the task graph to re-simulate state amplitudes incrementally.

\end{itemize}

We have evaluated the performance of qTask on a set of medium- and large-scale 
circuits in QASMBench~\cite{QASMBench}, 
an OpenQASM benchmark suite
for noisy intermediate-scale quantum (NISQ) evaluation and simulation.
Compared to two state-of-the-art simulators, Qulacs~\cite{Qulacs} and Qiskit~\cite{Qiskit},
qTask is respectively 1.46$\times$ and 1.71$\times$ faster for full simulation 
and 5.77$\times$ and 9.76$\times$ faster for incremental simulation.
We believe qTask stands out as a
unique system given the ensemble of software tradeoffs and
architectural decisions we have made.

\section{Background and Related Work}

In this section, we give an overview of quantum computation
and related work on QCS.
Then, we discuss the importance of incremental QCS and
its challenges.

\subsection{Quantum Circuits and Simulation}

A quantum circuit of $n$ qubits is a sequence of quantum gates that
act on \textit{quantum states}.
Each state $\psi$ is a \textit{superposition} or a linear combination of $2^n$
possible binary states using $2^n$ \textit{amplitudes},
denoted as
$\ket \psi = \alpha_{0} \ket{0...00} + \alpha_{1} \ket{0...01} + ... + \alpha_{2^n} \ket{1...11}$.
For brevity,
binary states can be written in decimal,
$\ket \psi = \alpha_{0} \ket{0} + \alpha_{1} \ket{1} + ... + \alpha_{2^n} \ket{2^n}$.
Squared amplitudes are probability of individual states
to which a superposition state will collapse
when measurement is performed.
Thus, squared amplitudes need to sum up to 1. 

\begin{equation*}
\sum_{b} |\alpha_b|^2 = 1
\end{equation*}

Industrial quantum computers use a set of \textit{standard} single-qubit gates 
and two-qubit controlled gates
to perform universal computation~\cite{QASMBench}.
These standard gates are defined by $2\times2$ or $4\times4$ unitary matrices
and can compose larger gates, such as Toffoli, Fredkin, and controlled rotators.
The following example shows the standard Pauli-X gate, Hadamard gate,
and controlled-NOT (CNOT) gate in matrix form.
Notice that NOT and X are interchangeable in gate naming.

\begin{equation*}
X =
\begin{bmatrix}
0 & 1 \\
1 & 0
\end{bmatrix},
H = \frac{1}{\sqrt{2}}
\begin{bmatrix}
1 & 1 \\
1 & -1
\end{bmatrix},
CX = 
\begin{bmatrix}
1 & 0 & 0 & 0 \\
0 & 1 & 0 & 0 \\
0 & 0 & 0 & 1 \\
0 & 0 & 1 & 0
\end{bmatrix}
\end{equation*}

\begin{figure}[!h]
  \centering
 \centerline{\includegraphics[width=1.\columnwidth]{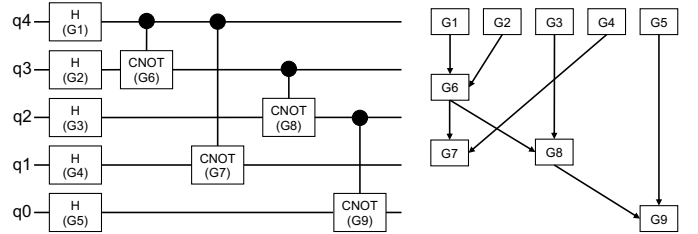}}
  \caption{
    A five-qubit quantum circuit of nine gates (left) and its gate dependency graph (right).
   }
  \label{fig::circuit_and_gate_dependency}
\end{figure}

A collection of quantum gates at a level forms a \textit{unitary transformation matrix}
defined by the Kronecker product ($\otimes$) 
of individual gate matrices from the first qubit to the $n$-th qubit.
Figure \ref{fig::circuit_and_gate_dependency} shows a five-qubit circuit
with five Hadamard gates and four CNOT gates.
The first five Hadamard gates form 
the $32\times32$ transformation matrix, $H^{\otimes5}$, to create superposition.
The last four CNOT gates create entanglement.
Finding unitary transformation matrices is an integral part of QCS.
First,
we order the gates left to right and pad an empty spot
with an identity matrix of an appropriate dimension.
Parallel gates can be ordered arbitrarily, 
for instance, \texttt{G7} and \texttt{G8}.
Then, we find all $2^n\times2^n$ matrices
via Kronecker product and multiply them in order.
The resulting matrix represents the entire circuit and can be multipled by input state vectors
to derive output states.
Such simulations allow researchers and developers to evaluate
the complexity of new quantum algorithms and validate quantum devices.

\subsection{Existing Quantum Circuit Simulators}

Mainstream QCS software is based on two paradigms,
\textit{state vector} and \textit{tensor network contraction}.
State vector-based QCS keeps a vector of the current state and iteratively
multiplies it by a state transformation matrix.
To improve space and time efficiency,
researchers have proposed various techniques,
such as compact binary decision diagram (BDD) to represent matrices~\cite{DDSim},
lossy data compression to trade accuracy for space~\cite{Wu_19_01},
multi-threaded sparse matrices~\cite{Qulacs},
graphics processing units (GPUs) to gain throughput performance~\cite{QGPU, cuQuantum},
and distributed vector to scale out computation~\cite{mpiQulacs, IntelQC}.
While being mathematically simple,
state vector has been widely used in mainstream simulators
including commercial tools (IBM Qiskit~\cite{Qiskit}, MS QDK~\cite{MSQDK}, Google Qsim~\cite{GoogleQSim}).

On the other hand,
tensor network-based QCS represents a quantum circuit in a tensor network
and explores the best contraction order for state update.
However, the time and space costs for contracting tensor networks 
are exponential with the network width.
Therefore, existing research has been targeting low-depth
circuits using various optimization techniques,
such as slicing window with asynchronous task parallelism~\cite{Jet, QuantumBoson},
GPU acceleration~\cite{cuQuantum}, 
and tree partitioning~\cite{OptimalTensorQCS}.
While computing tensor networks is efficient,
such an organization does not support intermediate measurement~\cite{Wu_19_01}.
Furthermore, tensor network is primarily optimized for static environments.
When a circuit begins to change,
maintaining a dynamic tensor network becomes very challenging.

In addition to state vector and tensor network,
general-purpose heuristics
for improving simulation efficiency have also been studied,
such as gate cancelling~\cite{FasterQCS}, gate restriction~\cite{CliffordGates}, 
gate reordering~\cite{QGPU}, 
pattern matching~\cite{PatternMatchingOpt}, 
approximation~\cite{ApproximatedQCS}, and so on.
Many of these strategies focus on removing
redundancy in a quantum circuit or restructuring it
to gain a more compact representation
under certain assumptions.

\subsection{Importance and Challenges of Incrementality}

As for the rapid growth of quantum software development,
\textit{incremental QCS} has emerged
as an important tool for the success of many simulation-driven QC applications~\cite{Gokhale_19_01}.
For instance, quantum circuit synthesis engines can issue thousands of simulation runs
in an optimization loop to evaluate how a local change 
(e.g., qubit swapping, rotation degree turning, gate insertion and removal) 
affects output amplitudes~\cite{DA4QC}.
This type of optimization is especially common
in cross-layer quantum computer designs that incrementally map software logic to hardware
with simulation in the loop~\cite{Classiq}.
When a small portion of a quantum circuit is modified, 
re-simulating the full circuit is infeasible 
from a turnaround time and productivity perspective.
The simulator must \textit{incrementally} update only affected regions
without exhaustive simulation.
The success of incremental QCS can also largely
improve the efficiency, and consequently user experience, 
of QC platforms that target interactive
learning of quantum algorithms with step-by-step simulation.
Unfortunately,
the current QCS landscape is largely short of incrementality.

\begin{figure}[!h]
  \centering
 \centerline{\includegraphics[width=1.\columnwidth]{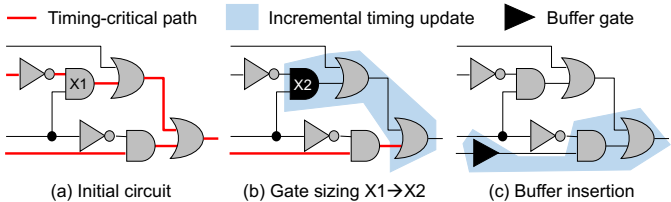}}
  \caption{
    Incremental timing analysis in the classical design flow of digital circuits~\cite{OpenTimerv2}.
    Optimization tools iteratively change the design and incrementally
    update timing information until no timing-critical paths are found.
   }
  \label{fig::classical_incrementality}
\end{figure}

On the contrary, 
incrementality has been extensively studied in the classical design flow 
of digital circuits.
For instance, design automation tools 
heavily count on incremental timing/power analysis algorithms
for efficient circuit optimization~\cite{Huang_14_01, OpenTimerv2}.
Figure \ref{fig::classical_incrementality} shows 
an example of timing-driven optimization.
These algorithms explore incrementality along the circuit network
and update quantities on a \textit{per-gate} basis
after optimization transforms (e.g., gate sizing, buffer insertion) change the design.
However, such ideas are not easily applicable to incremental QCS
because state values can be \textit{entangled} and therefore are
inseparable among gates.
This property also brings another challenge to parallelization.
For instance, although \texttt{G7} and \texttt{G8} in Figure \ref{fig::circuit_and_gate_dependency}
are structurally independent of each other,
we cannot apply these two CNOT operations simultaneously as race can occur
when both \texttt{q4} and \texttt{q3} are 1.
We need a different task decomposition strategy 
to parallelize incremental QCS.

Extending existing QCS algorithms to incorporate incrementality is nontrivial, either.
The biggest challenge is to maintain \textit{consistency}
or invariants in these algorithms when data structures of states and gates
start to change.
For instance, BDD can be extremely compact for full simulation~\cite{DDSim},
but its linear coupling between stages
can incur expensive reorganization of a BDD
when a local change happens in an early stage.
Likewise, algorithms that count on restructuring 
input circuits~\cite{PatternMatchingOpt} will need to manage an additional layer of consistency 
between the original and the modified circuits.
Similar challenges exist in other full simulation algorithms 
as well~\cite{Wu_19_01, FasterQCS}.
Although existing QCS algorithms and ideas in the classical design flow
have their benefits,
we believe a ground-up design of simulation system
is necessary to overcome the unique
challenges of incremental QCS.


\section{qTask: Task-parallel Incremental QCS}

In this section, we introduce \textit{qTask}, a new QCS engine 
that efficiently supports incrementality using task parallelism.
qTask introduces a lightweight C++ programming model 
for incremental simulation
and backs up the model with an efficient runtime that explores 
both inter- and intra-gate operation parallelisms from partitioned data blocks.
We first discuss the targeted environment of qTask
and then present its technical details.
Throughout the discussion,
we will use the circuit example in Figure \ref{fig::circuit_and_gate_dependency}
to explain key steps in qTask.

\subsection{Targeted Simulation Environment}

qTask targets medium-size \textit{gate-level} quantum circuits
on a single machine that has sufficient memory to store
all output amplitudes ($\alpha_i$).
While qTask currently assumes all data fit in memory,
the proposed partitioning strategy can be extended to
a higher number of qubits using out-of-core memory, which is part of our future work.
To comply with modern quantum computers,
we target \textit{standard gates} defined atomically
in OpenQASM~\cite{OpenQASM} and QASMBench~\cite{QASMBench}, 
as shown in Table \ref{tab::supported_standard_gates}.
These standard gates can be 1) mapped to machine-specific
gates for actual execution and 2)
assembled to form composition gates (e.g., CZ, CCX, SWAP).
Since qTask does not impose any constraints on gates,
composition gates can be also included to our database 
using the same simulation method as standard gates.

\begin{table}[!h]
\centering
\small
\caption{Supported standard quantum gates by qTask based on OpenQASM specification~\cite{OpenQASM}}
\begin{tabular}{|c|c|c|c|}
\hline
Gates & Functionality  & Gates & Functionality        \\\hline\hline
CNOT  & Controlled-NOT & SDG   & Conjugate of sqrt(Z) \\\hline
X     & Pauli-X gate   & T     & sqrt(S) phase        \\\hline
Y     & Pauli-Y gate   & TDG   & Conjugate of sqrt(S) \\\hline
Z     & Pauli-Z gate   & RX    & X-axis rotation      \\\hline
H     & Hadamard gate  & RY    & Y-axis rotation      \\\hline
S     & sqrt(Z) phase  & RZ    & Z-axis rotation      \\\hline
\end{tabular}
\label{tab::supported_standard_gates}
\end{table}

\subsection{Programming Model}

Unlike existing quantum programming models that do not anticipate incrementality,
qTask introduces a lightweight C++-based model with two new concepts:
First, qTask groups application programming interface (API) to three categories,
\textit{circuit modifier}, \textit{state update}, and \textit{query}.
The three categories describe operations that modify the circuit, 
update state amplitudes (incrementally), and query circuit quantities, respectively.
Second, qTask asks users to explicitly structure gates on a \textit{per-net} basis 
to facilitate the design of incremental QCS.
A net is a group of gates that are parallel in structure
(e.g., \texttt{G1}--\texttt{G5} in Figure \ref{fig::circuit_and_gate_dependency}).
Table \ref{tab::api} shows the key API to support incremental QCS in qTask.
Currently, qTask does not support adding or removing a qubit
as the number of qubits is typically decided in the beginning.

\begin{table}[!h]
\centering
\small
\caption{Key API to support incremental QCS in qTask}
\begin{tabular}{|c|c|}
\hline
Method & Functionality   \\\hline\hline
insert\_net     & insert a new empty net to the circuit \\\hline
remove\_net     & remove a net and all its gates from the circuit \\\hline
insert\_gate    & insert a net gate to an existing net \\\hline
remove\_gate    & remove a gate from its net and the circuit \\\hline
update\_state   & update the state value, incrementally \\\hline
dump\_graph     & dumps the current partition graph \\\hline
\end{tabular}
\label{tab::api}
\end{table}

Listing \ref{programming_model} shows an example of qTask code for 
creating the quantum circuit in Figure \ref{fig::circuit_and_gate_dependency}.
We start by creating a circuit object, \texttt{ckt}, with five qubits, 
\texttt{q4}, \texttt{q3}, \texttt{q2}, \texttt{q1}, and \texttt{q0},
where \texttt{q4} is the most significant bit.
Then, we create five nets using the method \texttt{insert\_net},
which inserts a new net right after the net given in the argument.
Since the five Hadamard gates are of no structural dependency,
we insert them to \texttt{net1}.
Next, we insert four CNOT gates to \texttt{net2}, \texttt{net3},
\texttt{net4}, and \texttt{net5}, respectively.
If a gate is inserted to a net that introduces a dependency,
such as \texttt{G6} and \texttt{G7},
qTask will throw an exception.
When we finish describing the circuit, 
calling \texttt{dump\_graph} will dump the current task graph 
of partitioned blocks to a DOT format.
Calling \texttt{update\_state} will run the task graph to perform simulation.
The last three lines modify the circuit by removing \texttt{G8} 
and inserting a new CNOT gate \texttt{G10}, followed by
calling \texttt{update\_state}
to re-simulate state amplitudes incrementally.

\begin{lstlisting}[language=C++,label=programming_model,caption={qTask code (before circuit modifiers) of Figure \ref{fig::circuit_and_gate_dependency}.}]
qTask ckt(5);
auto [q4, q3, q2, q1, q0] = ckt.qubits(); 
// create five nets and eight gates
auto net1 = ckt.insert_net(ckt.nets().begin());
auto net2 = ckt.insert_net(net1);
auto net3 = ckt.insert_net(net2);
auto net4 = ckt.insert_net(net3);
auto net5 = ckt.insert_net(net4);
auto G1 = ckt.insert_gate(H, net1, q4);
auto G2 = ckt.insert_gate(H, net1, q3);
auto G3 = ckt.insert_gate(H, net1, q2);
auto G4 = ckt.insert_gate(H, net1, q1);
auto G5 = ckt.insert_gate(H, net1, q0);
auto G6 = ckt.insert_gate(CNOT, net2, q3, q4);
auto G7 = ckt.insert_gate(CNOT, net3, q1, q4);
auto G8 = ckt.insert_gate(CNOT, net4, q2, q3);
auto G9 = ckt.insert_gate(CNOT, net5, q0, q2);
ckt.dump_graph(std::cout);
ckt.update_state();  // full update
// modify the circuit
ckt.remove_gate(G8);
auto G10 = ckt.insert_gate(CNOT, net4, q1, q2);
ckt.update_state();  // incremental update
\end{lstlisting}

Internally, qTask does not maintain any gate dependency graph, such as the one in Figure \ref{fig::circuit_and_gate_dependency},
but a list of nets based on the order of their constructions.
Since all the gates in a net are structurally parallel,
qTask can group them in arbitrary order to design an efficient 
memory management scheme atop our task partitioning, discussed later.

\subsection{Task Decomposition Strategy}

To facilitate the design of efficient incremental QCS,
qTask employs a top-down parallel decomposition strategy
using the \textit{task graph} model.
qTask divides a state vector into a set of disjoint, equal-size blocks
and groups consecutive blocks to form \textit{partitions}.
Each partition spawns one or multiple \textit{tasks} to perform
gate operations on designated memory regions.
This strategy
breaks down gate dependencies to \textit{task dependencies}
among partitions, enabling inter-gate operation parallelism.
If a partition contains more than one block,
qTask further spawns parallel tasks to
explore intra-gate operation parallelism among these blocks.
By leveraging existing task graph programming systems,
qTask transparently scales to many processors.
Here, we focus on task partitioning first and will discuss
how partitions are connected to each other as part of circuit modifiers and incremental update.

\begin{figure}[!h]
  \centering
 \centerline{\includegraphics[width=1.\columnwidth]{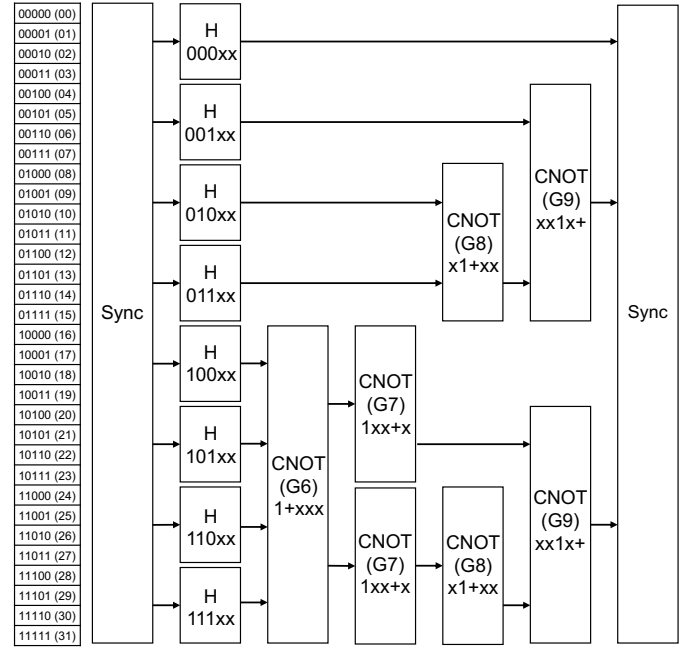}}
  \caption{
    Task partitioning for simulating the quantum circuit
    in Figure \ref{fig::circuit_and_gate_dependency}
    using a block size of 4.
    The task graph explores inter-gate parallelism through partitioned data blocks.
   }
  \label{fig::partition}
\end{figure}

Figure \ref{fig::partition} shows the partition diagram
for the simulation workload of the quantum circuit 
in Figure \ref{fig::circuit_and_gate_dependency}.
Each block size $B$ is a power of two (4 here)
and represents the minimum number of elements
or granularity for each task.
The key idea behind our partitioning is to 
carry out gate operations over a state vector in two modes,
\textit{non-superposition} and \textit{superposition} gates.
Gate operations, such as CNOT, diagonal matrices, and permutations
do not create superposition and can directly alter the state vector
using linear swapping and scaling.
For instance, the CNOT gate \texttt{G6} 
in Figure \ref{fig::circuit_and_gate_dependency}
is equivalent to swapping state amplitudes between
\texttt{10xxx} and \texttt{11xxx}, 
where ``\texttt{xxx}'' denotes all possible binary strings
of the first three qubits (\texttt{1+xxx} for short).  
On the other hand, gate operations that result in superposition, 
such as non-diagonal matrices and rotators,
will fall back to the use of state transformation matrix.

\begin{figure}[!h]
  \centering
 \centerline{\includegraphics[width=1.\columnwidth]{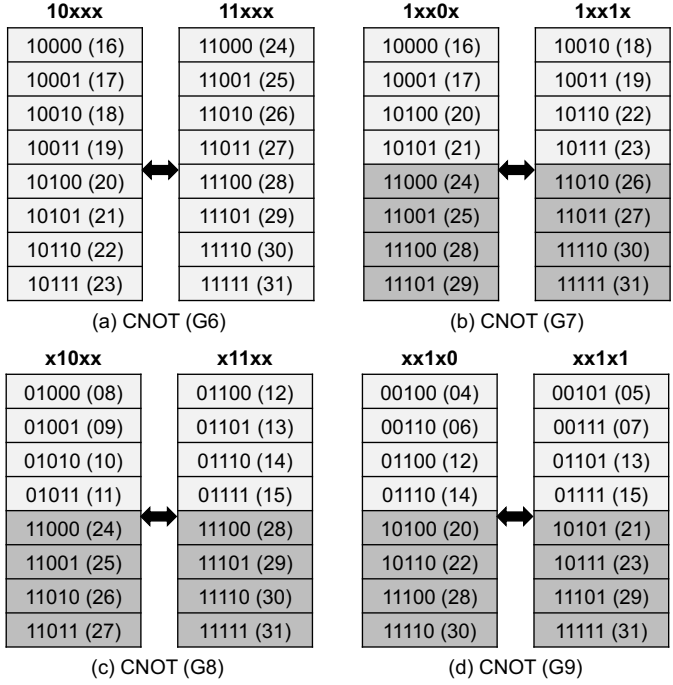}}
  \caption{
    Task partitioning of CNOT gates \texttt{G6}--\texttt{G9} in Figure \ref{fig::partition}. 
    (a) \texttt{G6} gives one partition spanning four consecutive data blocks.
    (b)-(c) Both \texttt{G7} and \texttt{G8} give two partitions 
    each spanning two consecutive data blocks.
    (d) \texttt{G9} gives two partitions each spanning three consecutive data blocks.
   }
  \label{fig::partition_c6_c7}
\end{figure}

\begin{figure}[!h]
  \centering
 \centerline{\includegraphics[width=1.\columnwidth]{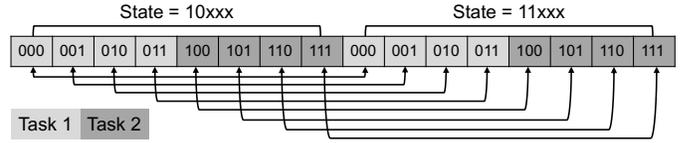}}
  \caption{
    Intra-gate operation parallelism of \texttt{G6} in Figure \ref{fig::partition_c6_c7}.
    Tasks 1 and 2 simultaneously operate on the first and the second blocks
    to swap amplitudes between states.
   }
  \label{fig::intra_gate_parallelism}
\end{figure}

Figure \ref{fig::partition_c6_c7} gives an example of 
how qTask performs CNOT using partitioned tasks.
For \texttt{G6}, we need to swap eight amplitudes
between \texttt{10xxx} and \texttt{11xxx}.
Since the block size is 4, the eight swaps can be parallelized
by two tasks, starting at states \texttt{10000} and \texttt{10100},
respectively.
However, the two tasks cannot appear in two parallel partitions
because their memory regions overlap 
(i.e., $[16, 27]$ and $[20, 31]$, using decimal representation).
Instead, we form one partition of $[16, 31]$,
as shown in Figure \ref{fig::partition},
and spawn the two parallel tasks within this partition 
to explore intra-gate operation parallelism,
as illustrated in Figure \ref{fig::intra_gate_parallelism}.
In qTask, intra-gate operation parallelism can be regarded as a parallel-for 
with chunk size equal to our block size.
On the other hand,
\texttt{G7} results in two partitions
that can run in parallel
because the memory regions of the two blocks are
$[16, 23]$ and $[24, 31]$ that do not overlap,
as shown in Figure \ref{fig::partition}.
The procedure to derive partitions for \texttt{G8} and \texttt{G9}
is similar.

The memory region of a block
can be quickly decided by 
replacing ``\texttt{x}''s with the binary string of a multiple of $B$.
In Figure \ref{fig::partition_c6_c7},
for instance,
the first and the second blocks of \texttt{G6} are the two states by
replacing \texttt{xxx} with \texttt{000} and \texttt{100}, respectively.
By iterating blocks in order,
we can decide when to form a partition of consecutive blocks
that overlap in memory regions.
Furthermore, it can be observed that the proposed partitioning
has a repetitive pattern (see Figure \ref{fig::partition}).
Once we have found a partition, the rest can be quickly
decided as they all have the same size by symmetry.
qTask employs this algorithm to decide partitions for gates
X, Y, Z, S, SDG, T, TDG, SWAP, and RX/RY/RZ of certain degrees 
that do not form superposition (e.g., RX($\pi$)).

For gate operations that form superposition (e.g., Hadamard, RX($\pi/2$)),
qTask falls back to the principle of state transformation matrix.
Since this process is equivalent to matrix-vector multiplication,
we partition the state vector into an equal number of blocks
and perform parallel multiplication.
For instance, the first net of five Hadamard gates 
in Figure \ref{fig::circuit_and_gate_dependency}
will result in eight partitions each of one block, 
as shown in Figure \ref{fig::partition}.
Each partition computes four amplitudes via multiplying
the input state vector by the corresponding subset of matrix rows.
Since the multiplication cannot start until the previous vector is ready,
the eight partitions are preceded by a synchronization task,
\texttt{sync}.
Notice that each partition derives its subset of matrix rows
on the fly to save memory and gain parallelism
using recursive tensor products. 
We stop the recursion when zero and identity patterns occur.

\subsection{Circuit Modifiers}

With a task graph in place,
we can efficiently perform incremental QCS
by restructuring the graph connectivity and 
identifying affected partitions to resimulate after a circuit modifier is applied.
Since qTask partitions a state vector into contiguous blocks,
connections between partitions can be quickly decided 
by a few forward and backward checks using range intersection algorithm.
Specifically, a connection exists between two partitions
if they are the closest pair of overlapped blocks.
By scanning neighboring partitions and their block coverages,
qTask can efficiently rebuild new connections and identify affected partitions
for incremental update.
We will focus on removing and inserting gates since
net-level circuit modifiers can be built on top.


\begin{figure}[!h]
  \centering
 \centerline{\includegraphics[width=1.\columnwidth]{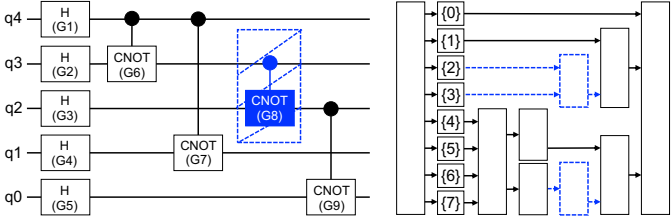}}
  \caption{
    Remove gate \texttt{G8} from the quantum circuit 
    in Figure \ref{fig::circuit_and_gate_dependency}
    and its impact on partitioned data blocks. 
    Numbers in ``\{\}'' denote block IDs.
   }
  \label{fig::remove_gate}
\end{figure}

Figure \ref{fig::remove_gate} illustrates how removing a gate affects the graph connectivity.
When we remove gate \texttt{G8},
all its partitions and relevant dependencies 
should be removed (marked in blue dash).
For each removed partition,
we need to reconnect its preceding partitions
to its successor partitions if an overlap exist in their blocks.
Since each partition is a group of consecutive blocks,
by keeping a list of block indices covered by each partition,
we can quickly decide overlapped partitions using
range intersection algorithm.
For instance, the first partition of \texttt{G8} spans
the block range $[2, 3]$, which intersects the block range $[1, 3]$
of its successor.

\begin{figure}[!h]
  \centering
 \centerline{\includegraphics[width=1.\columnwidth]{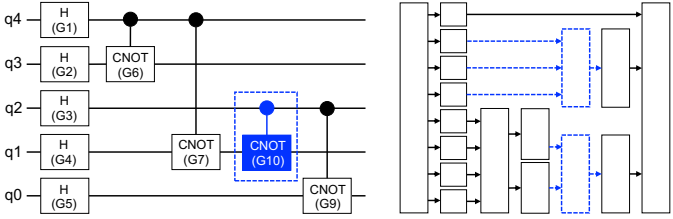}}
  \caption{
    Insert a new gate \texttt{G10} after the removal of \texttt{G8}
    in Figure \ref{fig::remove_gate}
    and its impact on partitioned data blocks.
   }
  \label{fig::insert_gate}
\end{figure}

\begin{figure}[!h]
  \centering
 \centerline{\includegraphics[width=1.\columnwidth]{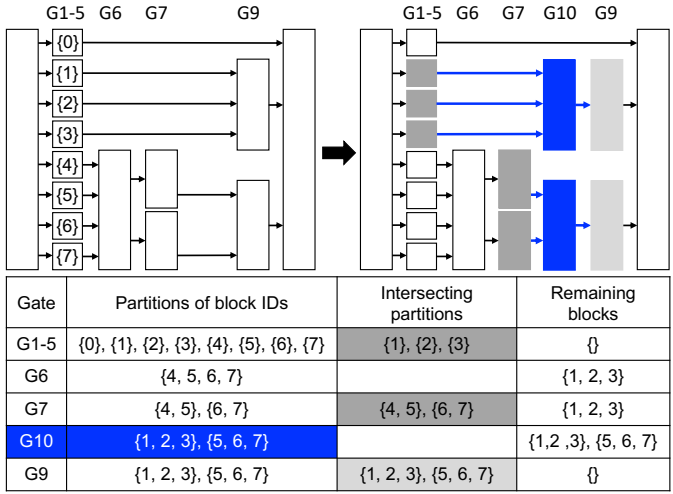}}
  \caption{
    Incremental update of the task graph 
    after inserting a new gate \texttt{G10} to Figure \ref{fig::remove_gate}.
   }
  \label{fig::incremental_update_graph}
\end{figure}

Figure \ref{fig::insert_gate} illustrates how inserting a new gate affects the graph connectivity.
When a new gate \texttt{G10} is inserted,
we first identify its partitions (marked in blue dash) and then connect each partition
with appropriate predecessors and successors.
Figure \ref{fig::incremental_update_graph} illustrates our algorithm
to find such predecessors and successors.
Starting from the row of \texttt{G10}, which has two partitions of block ranges
$[1, 3]$ and $[5, 7]$,
we iteratively move backward and forward to find intersected partitions 
for predecessors and successors
until the remaining blocks of \texttt{G10} become empty.
For example,
with one step forward, the two partitions of \texttt{G9} can all cover that of \texttt{G10},
resulting in two successor dependencies.
Similarly,
with one step backward, the two partitions of \texttt{G7} cover only the second partition of \texttt{G10},
resulting in two predecessor dependencies;
with two more steps backward, we have three predecessor dependencies that completely cover the first partition of \texttt{G10}.
Since dependency constraints are transitive,
we remove existing dependencies between these predecessors and successors.

%



\subsection{Incremental Update}

qTask keeps a list of partitions called \textit{frontiers} for each sequence of circuit modifiers.
Frontiers are the source to start incremental update of affected state amplitudes
when users issue an update call after a sequence of circuit modifiers.
For each newly inserted gate, we add all its partitions to the frontier list.
For each removed gate, we add all successors of removed partitions to the frontier list.
Now, it should be clear that our partitioning strategy effectively 
scopes down state update to only successor partitions 
that are reachable from frontiers.
Such successors can be found through a depth-first-search (DFS) 
starting from each frontier partition.

\begin{figure}[!h]
  \centering
 \centerline{\includegraphics[width=1.\columnwidth]{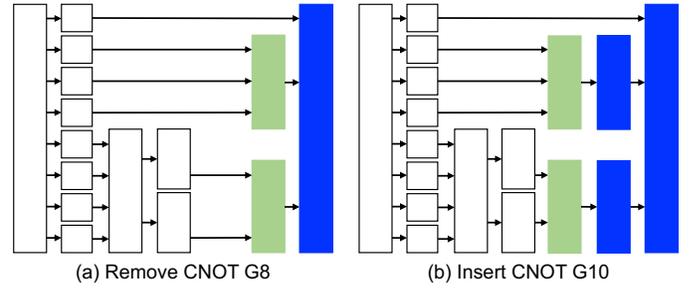}}
  \caption{
    Incremental update of state amplitudes 
    after (a) removing \texttt{G8} as in Figure \ref{fig::remove_gate}
    and (b) inserting \texttt{G10} as in Figure \ref{fig::insert_gate}.
    Frontier partitions are marked in green, and their reachable successors are marked in blue.
   }
  \label{fig::incremental_update}
\end{figure}

\begin{figure}[!h]
  \centering
 \centerline{\includegraphics[width=1.\columnwidth]{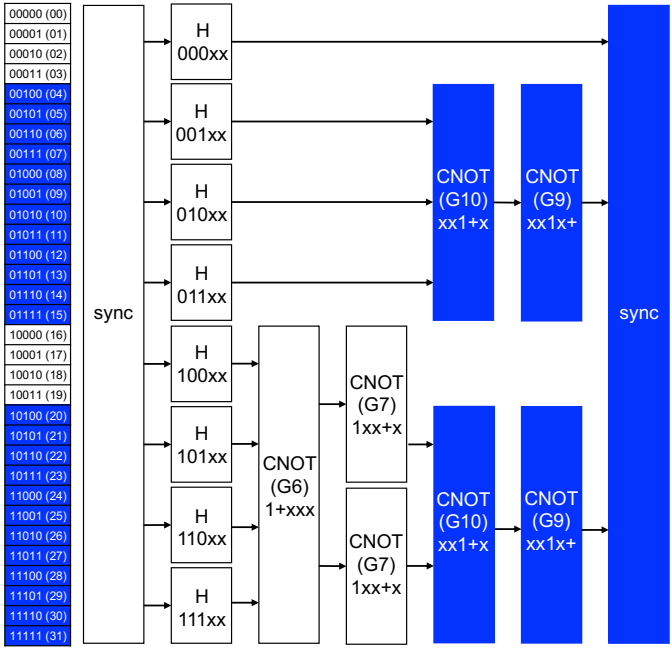}}
  \caption{
   Detailed state map of Figure \ref{fig::incremental_update}(b),
   in which only four partitions (marked in blue) are involved in 
   updating 24 state amplitudes incrementally.
   }
  \label{fig::incremental_update_details}
\end{figure}

Figure \ref{fig::incremental_update} illustrates how qTask identifies frontiers and
use them to carry out incremental update for
Figure \ref{fig::remove_gate} and Figure \ref{fig::insert_gate}.
In (a), the two successor partitions of \texttt{G8} (marked in green)
are frontiers when \texttt{G8} is removed from the circuit.
Intuitively speaking, we only need to update state amplitudes of the two partitions 
and onward since removing \texttt{G8} has no impact on other partitions.
Similarly in (b), inserting \texttt{G10} to the circuit 
introduces two new partitions to perform CNOT operations.
The two partitions will be added to the frontier list and all their successors 
will participate in the incremental update.
Figure \ref{fig::incremental_update_details}
shows a detailed state map of Figure \ref{fig::incremental_update}(b).
We can see only 24 state amplitudes ($[4, 15]$ and $[20, 31]$) are incrementally updated 
after removing \texttt{G8} and inserting \texttt{G10}.

\subsection{Implementation Details}

In this section, we discuss three important implementation details
of qTask, \textit{task graph creation}, \textit{per-net state vector management},
and \textit{copy-on-write data optimization}.

\subsubsection{Task graph creation}

\begin{figure}[!h]
  \centering
 \centerline{\includegraphics[width=1.\columnwidth]{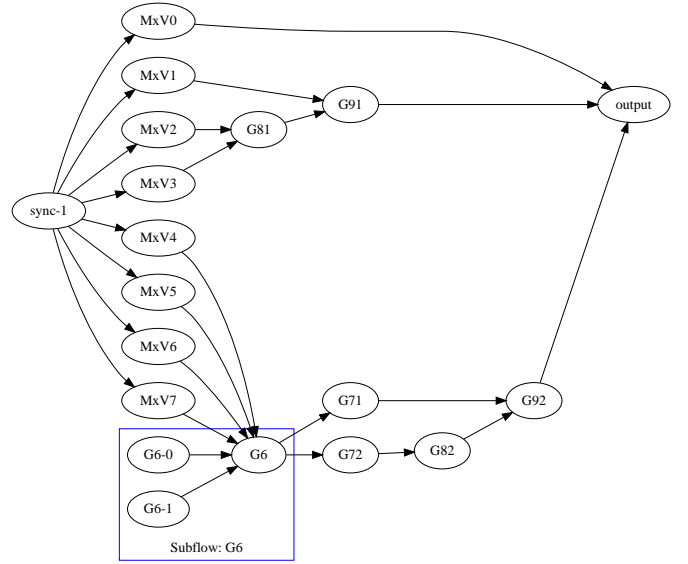}}
  \caption{
   Taskflow graph of Figure \ref{fig::partition}.
   Each task performs a specific gate operation on partitioned
   data blocks.
   The subflow task, \texttt{G6} (in blue box), spawns two static tasks 
   for intra-gate operation parallelism
   as described in Figure \ref{fig::intra_gate_parallelism}.
   }
  \label{fig::taskflow}
\end{figure}

We leverage the Taskflow library~\cite{taskflow}
to derive a task graph from partitioned data blocks.
We decide to use Taskflow because of its simplicity 
and many successful use cases in classical circuit design~\cite{Huang_19_01, Huang_19_02, Huang_20_03, Huang_21_01, Huang_21_02, RTLflow, GenFuzz, HeteroCPPR, Chiu_22_01, Zhou_22_01, Chiu_22_02, Guo_23_01},
but other tasking libraries (e.g., TBB~\cite{TBBBook}, OpenMP~\cite{OpenMP}) are also possible.
Specifically, we use the static tasking and dynamic tasking (subflow) features~\cite{Lin_19_01}
of Taskflow to compose inter- and intra-gate operation parallelisms,
respectively.
Figure \ref{fig::taskflow} shows the Taskflow graph
of Figure \ref{fig::partition},
where 1) 16 static tasks 
(\texttt{sync-1}, \texttt{MxV0}--\texttt{MxV7}, \texttt{G71}, \texttt{G72}, 
\texttt{G81}, \texttt{G82}, \texttt{G91}, \texttt{G92}, \texttt{output})
are used to formulate inter-gate operation parallelism 
and 2) one subflow of two static tasks 
(\texttt{G6-0} and \texttt{G6-1})
is used to formulate intra-gate operation parallelism of \texttt{G6}.
Each time the application requests a state update,
qTask derives a new Taskflow graph from affected partitions
and submits it to Taskflow's work-stealing runtime~\cite{Lin_20_01}
for parallel incremental simulation.

\subsubsection{Per-net state vector management}
Since gates in the same net are structurally independent of each other,
we first group superposition gates to share a state vector and partition it to an equal
number of blocks,
such as \texttt{MxV0}--\texttt{MxV7} in Figure \ref{fig::taskflow}.
These partitions will succeed an empty task 
(\texttt{sync-1})
that synchronizes all previous tasks
to safely perform parallel matrix-vector multiplication.
Next, we create a state vector for each non-superposition gate in the net
and partition the vector into a set of consecutive blocks based on the proposed method.
If multiple gates exist, we connect them in an increasing order of block count in partitions.
This is because a partition of a high block count tends 
to affect a large number of downstream partitions, 
and we prefer to defer it as much as possible.

\subsubsection{Copy-on-write data optimization}
While qTask keeps multiple state vectors per net to store intermediate results for incrementality,
each state vector does not explicitly store all partitioned blocks.
Instead, we leverage the \textit{copy-on-write (COW)} technique~\cite{COW}
to optimize data access.
Each block has a COW C++ smart pointer to its predecessor block.
The memory and data of a block will not be created and copied until a task performs gate operations
on the block.
This COW optimization has two significant advantages:
First, qTask will only create necessary data blocks for simulation.
For example, the first and the fifth blocks of \texttt{G9} in Figure \ref{fig::partition}
will not be created unless explicitly requested.
Second,
multiple memory allocations and data copies between blocks can be simultaneously 
performed through parallel tasks.


\section{Experimental Results}

\begin{figure}[!h]
  \centering
 \centerline{\includegraphics[width=1.\columnwidth]{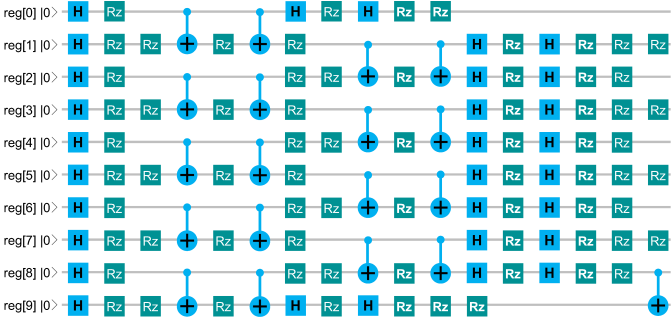}}
  \caption{
   Quantum circuit (partial) for Ising model simulation.
   }
  \label{fig::ising_n26}
\end{figure}

We evaluated the performance of qTask on 20 quantum circuits selected from
medium- and large-scale QASMBench~\cite{QASMBench}.
As shown in Table \ref{tab::overall_results},
these circuits exhibit different complexities in terms of
numbers of qubits and standard gates.
For example, vqe\_uccsd has the largest gate count of 10808, and 
big\_ising has the largest qubit count of 26.
All circuits except bb84 incorporate several CNOT gates to 
entangle and disentangle states.
Figure \ref{fig::ising_n26} shows a fraction
of the circuit, ising, that performs Ising model simulation
using 10 qubits.
All experiments ran on a CentOS 64-bit machine with 16 Intel i7 cores
at 2.50 GHz and 128 GB RAM.
We compiled qTask using clang++12
with C++17 standard \texttt{-std=c++17} 
and optimization flag \texttt{-O3} enabled.
The default block size of qTask is 256.
All data is an average of 10 runs.

\begin{table*}[!h]
\small
\centering
\caption{Performance comparison of qTask with Qulacs and Qiskit on medium- and large-scale QASMBench circuits~\cite{QASMBench}}
\resizebox{\textwidth}{!}{\begin{tabular}{|c|c|c|c|c|ccc|ccc|ccc|}
\hline
\multirow{2}{*}{Circuit} & \multirow{2}{*}{Description}    & \multirow{2}{*}{Qubits} & \multirow{2}{*}{Gates} & \multirow{2}{*}{CNOT} & \multicolumn{3}{c|}{Qulacs}                                                                                                                                                                                   & \multicolumn{3}{c|}{Qiskit}                                                                                                                                                                             & \multicolumn{3}{c|}{qTask}                                                                                                                                                                              \\ \cline{6-14}
                         &                                 &                         &                        &                       & \multicolumn{1}{c|}{\begin{tabular}[c]{@{}c@{}}full\\ (ms)\end{tabular}} & \multicolumn{1}{c|}{\begin{tabular}[c]{@{}c@{}}inc\\ (ms)\end{tabular}} & \begin{tabular}[c]{@{}c@{}}mem\\ (GB)\end{tabular} & \multicolumn{1}{c|}{\begin{tabular}[c]{@{}c@{}}full\\ (ms)\end{tabular}} & \multicolumn{1}{c|}{\begin{tabular}[c]{@{}c@{}}inc\\ (ms)\end{tabular}} & \begin{tabular}[c]{@{}c@{}}mem\\ (GB)\end{tabular} & \multicolumn{1}{c|}{\begin{tabular}[c]{@{}c@{}}full\\ (ms)\end{tabular}} & \multicolumn{1}{c|}{\begin{tabular}[c]{@{}c@{}}inc\\ (ms)\end{tabular}} & \begin{tabular}[c]{@{}c@{}}mem\\ (GB)\end{tabular} \\ \hline\hline
dnn                      & Quantum deep neural network     & 8                       & 1200                   & 384                   & \multicolumn{1}{c|}{21.8}                                                & \multicolumn{1}{c|}{2167.8}                                             & 0.07                                               & \multicolumn{1}{c|}{51.4}                                                & \multicolumn{1}{c|}{5114.3}                                             & 0.07                                               & \multicolumn{1}{c|}{22.4}                                                & \multicolumn{1}{c|}{529.3}                                               & 0.09                                               \\ \hline
adder                    & Quantum ripple adder            & 10                      & 142                    & 65                    & \multicolumn{1}{c|}{17.2}                                                & \multicolumn{1}{c|}{186.4}                                              & 0.05                                               & \multicolumn{1}{c|}{29.5}                                                & \multicolumn{1}{c|}{320.1}                                              & 0.04                                               & \multicolumn{1}{c|}{11.79}                                               & \multicolumn{1}{c|}{57.9}                                               & 0.06                                               \\ \hline
bb84                     & Quantum key distribution        & 8                       & 27                     & 0                     & \multicolumn{1}{c|}{1.1}                                                 & \multicolumn{1}{c|}{2.3}                                                & 0.03                                               & \multicolumn{1}{c|}{1.1}                                                 & \multicolumn{1}{c|}{2.4}                                                & 0.03                                               & \multicolumn{1}{c|}{1.5}                                                 & \multicolumn{1}{c|}{1.9}                                                & 0.04                                               \\ \hline
bv                       & Berstein-Vazirani algorithm     & 14                      & 41                     & 13                    & \multicolumn{1}{c|}{9.0}                                                 & \multicolumn{1}{c|}{21.7}                                               & 0.11                                               & \multicolumn{1}{c|}{16.7}                                                & \multicolumn{1}{c|}{40.6}                                               & 0.12                                               & \multicolumn{1}{c|}{6.7}                                                 & \multicolumn{1}{c|}{14.3}                                                & 0.13                                               \\ \hline
ising                    & Ising model simulation          & 10                      & 480                    & 90                    & \multicolumn{1}{c|}{49.6}                                                & \multicolumn{1}{c|}{1438.1}                                             & 0.08                                               & \multicolumn{1}{c|}{81.4}                                                & \multicolumn{1}{c|}{2360.1}                                             & 0.09                                               & \multicolumn{1}{c|}{41.7}                                                & \multicolumn{1}{c|}{550.14}                                              & 0.10                                               \\ \hline
multiplier               & Quantum multiplication           & 15                      & 574                    & 246                   & \multicolumn{1}{c|}{150.9}                                               & \multicolumn{1}{c|}{4199.0}                                             & 1.98                                               & \multicolumn{1}{c|}{283.7}                                               & \multicolumn{1}{c|}{7896.3}                                             & 2.86                                               & \multicolumn{1}{c|}{101.62}                                              & \multicolumn{1}{c|}{1052.6}                                              & 3.46                                               \\ \hline
multiplier\_35           & 3$\times$5 matrix multiplication       & 13                      & 98                     & 40                    & \multicolumn{1}{c|}{22.4}                                                & \multicolumn{1}{c|}{130.1}                                              & 0.10                                               & \multicolumn{1}{c|}{47.1}                                                & \multicolumn{1}{c|}{273.54}                                             & 0.15                                               & \multicolumn{1}{c|}{16.01}                                               & \multicolumn{1}{c|}{92.7}                                               & 0.18                                               \\ \hline
qaoa                     & Approximation optimization      & 6                       & 270                    & 54                    & \multicolumn{1}{c|}{5.4}                                                 & \multicolumn{1}{c|}{148.5}                                              & 0.01                                               & \multicolumn{1}{c|}{13.4}                                                & \multicolumn{1}{c|}{368.5}                                              & 0.01                                               & \multicolumn{1}{c|}{6.1}                                                 & \multicolumn{1}{c|}{37.65}                                               & 0.02                                               \\ \hline
qf21                     & Quantum factorization of 21     & 15                      & 311                    & 115                   & \multicolumn{1}{c|}{79.8}                                                & \multicolumn{1}{c|}{1173.1}                                             & 1.59                                               & \multicolumn{1}{c|}{191.5}                                               & \multicolumn{1}{c|}{2815.1}                                             & 1.66                                               & \multicolumn{1}{c|}{58.3}                                                & \multicolumn{1}{c|}{480.7}                                               & 1.91                                               \\ \hline
qft                      & Quantum Fourier transform       & 15                      & 540                    & 210                   & \multicolumn{1}{c|}{142.0}                                               & \multicolumn{1}{c|}{3621.0}                                             & 2.75                                               & \multicolumn{1}{c|}{281.2}                                               & \multicolumn{1}{c|}{7170.1}                                             & 3.11                                               & \multicolumn{1}{c|}{102.2}                                               & \multicolumn{1}{c|}{949.4}                                              & 3.17                                               \\ \hline
qpe                      & Quantum phase estimation        & 9                       & 123                    & 43                    & \multicolumn{1}{c|}{10.3}                                                & \multicolumn{1}{c|}{100.42}                                             & 0.02                                               & \multicolumn{1}{c|}{27.8}                                                & \multicolumn{1}{c|}{270.4}                                              & 0.04                                               & \multicolumn{1}{c|}{7.65}                                                & \multicolumn{1}{c|}{80.44}                                              & 0.05                                               \\ \hline
sat                      & Boolean satisfiability solver   & 11                      & 679                    & 252                   & \multicolumn{1}{c|}{85.5}                                                & \multicolumn{1}{c|}{3660.7}                                             & 0.11                                               & \multicolumn{1}{c|}{196.7}                                               & \multicolumn{1}{c|}{8422.1}                                             & 0.21                                               & \multicolumn{1}{c|}{62.3}                                                & \multicolumn{1}{c|}{786.5}                                               & 0.28                                               \\ \hline
seca                     & Shor's algorithm                & 11                      & 216                    & 84                    & \multicolumn{1}{c|}{28.4}                                                & \multicolumn{1}{c|}{401.0}                                              & 0.06                                               & \multicolumn{1}{c|}{59.64}                                               & \multicolumn{1}{c|}{843.0}                                              & 0.09                                               & \multicolumn{1}{c|}{21.42}                                               & \multicolumn{1}{c|}{128.5}                                               & 0.11                                               \\ \hline
simons                   & Simon's algorithm               & 6                       & 44                     & 14                    & \multicolumn{1}{c|}{0.83}                                                & \multicolumn{1}{c|}{3.9}                                                & 0.03                                               & \multicolumn{1}{c|}{1.44}                                                & \multicolumn{1}{c|}{6.71}                                               & 0.03                                               & \multicolumn{1}{c|}{0.81}                                                & \multicolumn{1}{c|}{2.44}                                               & 0.04                                               \\ \hline
vqe\_uccsd               & Variational quantum eigensolver & 8                       & 10808                  & 5488                  & \multicolumn{1}{c|}{244.4}                                               & \multicolumn{1}{c|}{249084.2}                                           & 0.36                                               & \multicolumn{1}{c|}{435.1}                                               & \multicolumn{1}{c|}{443367.1}                                           & 0.56                                               & \multicolumn{1}{c|}{259.4}                                               & \multicolumn{1}{c|}{44251.1}                                              & 0.76                                               \\ \hline
big\_adder               & Quantum ripple adder            & 18                      & 284                    & 130                   & \multicolumn{1}{c|}{200.1}                                               & \multicolumn{1}{c|}{2401.3}                                             & 7.98                                               & \multicolumn{1}{c|}{360.4}                                               & \multicolumn{1}{c|}{4300.8}                                             & 11.4                                               & \multicolumn{1}{c|}{137.9}                                               & \multicolumn{1}{c|}{602.5}                                              & 13.9                                               \\ \hline
big\_bv                  & Bernstein-Vazirani algorithm     & 19                      & 56                     & 18                    & \multicolumn{1}{c|}{125.0}                                               & \multicolumn{1}{c|}{305.9}                                              & 2.6                                                & \multicolumn{1}{c|}{234.5}                                               & \multicolumn{1}{c|}{573.9}                                              & 3.9                                                & \multicolumn{1}{c|}{95.4}                                                & \multicolumn{1}{c|}{126.6}                                              & 4.9                                                \\ \hline
big\_cc                  & Counterfeit coin finding        & 18                      & 34                     & 17                    & \multicolumn{1}{c|}{24.9}                                                & \multicolumn{1}{c|}{47.8}                                               & 0.98                                               & \multicolumn{1}{c|}{42.3}                                                & \multicolumn{1}{c|}{63.3}                                               & 1.5                                                & \multicolumn{1}{c|}{16.6}                                                & \multicolumn{1}{c|}{24.5}                                               & 1.7                                                \\ \hline
big\_ising               & Ising model simulation          & 26                      & 280                     & 50                    & \multicolumn{1}{c|}{1939.1}                                              & \multicolumn{1}{c|}{3345.5}                                             & 89.4                                               & \multicolumn{1}{c|}{1745.3}                                              & \multicolumn{1}{c|}{2866.2}                                             & 91.4                                               & \multicolumn{1}{c|}{991.4}                                               & \multicolumn{1}{c|}{2000.3}                                             & 114.3                                              \\ \hline
big\_qft                 & Quantum Fourier transform       & 20                      & 970                    & 380                   & \multicolumn{1}{c|}{2936.3}                                              & \multicolumn{1}{c|}{100567.0}                                           & 67.3                                               & \multicolumn{1}{c|}{3012.6}                                              & \multicolumn{1}{c|}{144453.4}                                             & 77.6                                               & \multicolumn{1}{c|}{2209.7}                                              & \multicolumn{1}{c|}{12912.8}                                             & 91.2                                               \\ \hline
\multicolumn{1}{l}{}                           & \multicolumn{1}{l}{}                                 & \multicolumn{1}{l}{}                         & \multicolumn{1}{l}{}                        & \multicolumn{1}{l|}{} & \multicolumn{1}{c|}{\textbf{1.46}}                                       & \multicolumn{1}{c|}{\textbf{5.77}}                                       & \textbf{0.74}                                       & \multicolumn{1}{c|}{\textbf{1.71}}                                       & \multicolumn{1}{c|}{\textbf{9.76}}                                       & \textbf{0.82}                                       & \multicolumn{1}{c|}{\textbf{1.00}}                                        & \multicolumn{1}{c|}{\textbf{1.00}}                                       & \textbf{1.00}                                       \\ \cline{6-14}
\end{tabular}}
\label{tab::overall_results}
\begin{tablenotes}[flushleft]
\item [1]\textbf{Qubits}: number of qubits \ \ \ \ \ \textbf{Gates}: number of standard gates \ \ \ \ \ \textbf{CNOT}: number of CNOT gates to entangle and disentangle states
\item [2]\textbf{full}: runtime of full simulation \ \ \ \ \ \textbf{inc}: runtime of incremental simulation \ \ \ \ \ \textbf{mem}: maximum resident set size (RSS)
\end{tablenotes}
\end{table*}

\subsection{Baseline}

Given the large number of quantum circuit simulators,
it is impractical to compare qTask with all of them.
Instead, we consider Qulacs~\cite{Qulacs} and Qiskit~\cite{Qiskit} as the baseline 
for the following three reasons:
First, both Qulacs and Qiskit have an optimized C++ back-end
and have demonstrated superior runtime performance over existing simulators~\cite{Qulacs}.
Second, Qulacs is completely open-source and 
has relatively rich documentation for C++ in addition to Python,
allowing us to reason the source when incrementality is taken into account.
Third, Qulacs and Qiskit support circuit modification,
in spite of no incremental simulation.
For example, Qulacs have API for inserting/removing gates at given positions, while
Qiskit allows adding/erasing gates as a byproduct of Python list.

\subsection{Overall Simulation Performance}

Table \ref{tab::overall_results} presents the overall simulation performance of qTask,
Qulacs, and Qiskit, using a maximum hardware concurrency of 16 threads.
In terms of full simulation, which issues a simulation call when the entire circuit is constructed,
qTask outperforms Qulacs and Qiskit in nearly all circuits.
The average speed-ups of qTask over Qulacs and Qiskit across all circuits
are $1.46\times$ and $1.71\times$, respectively.
We attribute this result to our partitioning strategy
which explores both inter- and intra-gate operation parallelisms.
This performance advantage becomes even more significant at larger circuits
that produce more partitioned tasks and parallelism than small ones.
For example,
qTask simulates big\_ising (26 qubits) 1.67$\times$ and 1.43$\times$ faster than Qulacs and Qiskit.
For circuits of state sizes below our partition size (i.e., 8 qubits with 256 amplitudes),
such as dnn, bb84, qaoa, and vqe\_uccsd
qTask is a bit slower than Qulacs because there is no much task parallelism.
Yet, the difference is very negligible (e.g., about 5\% in vqe\_uccsd).

Next, we study the performance of incremental simulation.
Following the convention of QASMBench,
we create a net per level and insert all parallel gates at that level to the net.
Starting from the first level, we repeat this process and issue level-by-level simulation calls
until the entire circuit is constructed.
That is, the number of simulation calls is equal to the number of nets or the circuit level/depth.
With incremental simulation, we clearly see the advantage of qTask.
On average, qTask is 5.77$\times$ and 9.76$\times$ faster
than Qulacs and Qiskit, respectively.
When a circuit include many gates in a long depth,
this advantage becomes even more pronounced.
Taking big\_qft for example, 
qTask finished 7.79$\times$ and 11.19$\times$ faster
than Qulacs and Qiskit, respectively.
The trade-off of this large performance gain is higher memory usage,
since we keep several state vectors per net to store intermediate results for
incrementality.
On average, qTask is 26\% and 18\% higher than Qulacs and Qiskit,
both of which incorporate specialized sparse data structures for state vectors and matrices.
However, how to efficiently extend such sparsity management to an incremental environment 
remains unknown.

\subsection{Performance of Incremental Simulation}

We further study the performance difference of incremental simulation 
between qTask and Qulacs over different numbers of circuit modifiers.
Hereafter, we compare with only Qulacs since Qiskit is much slower.
We follow the convention of classical design flow~\cite{OpenTimerv2, Huang_15_01, Huang_20_02, Huang_21_03} to 
define one \textit{incremental iteration} as a sequence of circuit modifiers
followed by a simulation call.
We first demonstrate the simulation performance for incremental gate insertions.
At each incremental iteration, we randomly select a few levels and insert all their gates
into the circuit.
Then, we call state update to re-simulate the modified circuit.
Iterations stop until the circuit is fully constructed.
Figure \ref{fig::runtime_incremental_gate_insertions}
draws the cumulative runtime over all incremental iterations
for two circuits, qft and big\_adder.
As the number of incremental iterations increases, 
the performance gap between qTask and Qulacs becomes larger.

\begin{figure}[!h]
  \centering
  \pgfplotsset{
    title style={font=\LARGE},
    label style={font=\LARGE},
    legend style={font=\large},
  }
  \begin{tikzpicture}[scale=0.50]
    \begin{axis}[
      title=qft,
      ylabel=Cumulative runtime (ms),
      xlabel=Incremental iterations,
      legend pos=north west,
    ]
    \addplot+ table[x=iter,y=qtask,col sep=space]{Fig/runtime_incremental_iterations_qft.txt};
    \addplot+ table[x=iter,y=qulacs,col sep=space]{Fig/runtime_incremental_iterations_qft.txt};
    \legend{qTask, Qulacs}
    \end{axis}
  \end{tikzpicture}
  \begin{tikzpicture}[scale=0.50]
    \begin{axis}[
      title=big\_adder,
      ylabel=Cumulative runtime (ms),
      xlabel=Incremental iterations,
      legend pos=north west,
    ]
    \addplot+ table[x=iter,y=qtask,col sep=space]{Fig/runtime_incremental_iterations_big_adder.txt};
    \addplot+ table[x=iter,y=qulacs,col sep=space]{Fig/runtime_incremental_iterations_big_adder.txt};
    \legend{qTask, Qulacs}
    \end{axis}
  \end{tikzpicture}
  \caption{Performance of incremental simulation for random gate insertions on two quantum circuits, qft and big\_adder.}
  \label{fig::runtime_incremental_gate_insertions}
\end{figure}
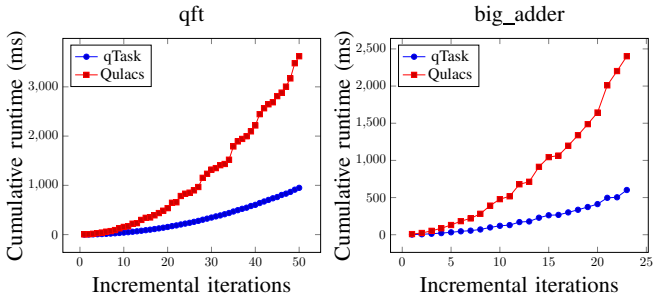
\begin{figure}[!h]
  \centering
  \pgfplotsset{
    title style={font=\LARGE},
    label style={font=\LARGE},
    legend style={font=\large},
  }
  \begin{tikzpicture}[scale=0.50]
    \begin{axis}[
      title=qft,
      ylabel=Runtime (ms),
      xlabel=Incremental iterations,
      legend pos=north east,
    ]
    \addplot+ table[x=iter,y=qtask,col sep=space]{Fig/runtime_incremental_removals_qft.txt};
    \addplot+ table[x=iter,y=qulacs,col sep=space]{Fig/runtime_incremental_removals_qft.txt};
    \legend{qTask, Qulacs}
    \end{axis}
  \end{tikzpicture}
  \begin{tikzpicture}[scale=0.50]
    \begin{axis}[
      title=big\_adder,
      ylabel=Runtime (ms),
      xlabel=Incremental iterations,
      legend pos=north east,
    ]
    \addplot+ table[x=iter,y=qtask,col sep=space]{Fig/runtime_incremental_removals_big_adder.txt};
    \addplot+ table[x=iter,y=qulacs,col sep=space]{Fig/runtime_incremental_removals_big_adder.txt};
    \legend{qTask, Qulacs}
    \end{axis}
  \end{tikzpicture}
  \caption{Performance of incremental simulation for random gate removals on two quantum circuits, qft and big\_adder.}
  \label{fig::runtime_incremental_gate_removals}
\end{figure}
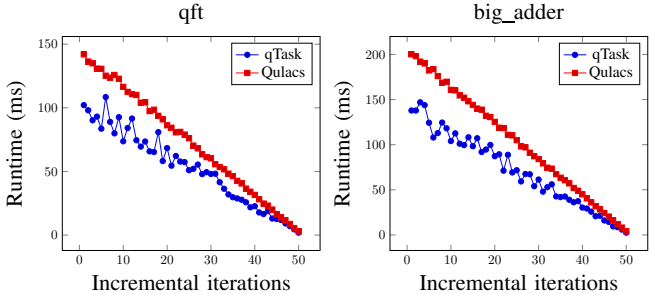

Next, we demonstrate the simulation performance for incremental gate removals.
Starting from a complete circuit, each incremental iteration
randomly selects a few levels and removes all their gates from the circuit.
Then, we call state update to re-simulate the modified circuit.
Iterations stop until the circuit becomes empty.
Figure \ref{fig::runtime_incremental_gate_removals}
draws the runtime at each incremental iteration
for the same circuits, qft and big\_adder.
Notice that the runtime at the iteration 0 represents full simulation.
As the number of incremental iterations increases,
both Qulacs and qTask approach zero due to fewer gates to re-simulate,
but qTask is always faster.
qTask fluctuates more than Qulacs 
because the number of affected partitions varies
across different incremental iterations.
Removing gates at a later level will affect fewer downstream partitions
that an earlier level, and vice versa.

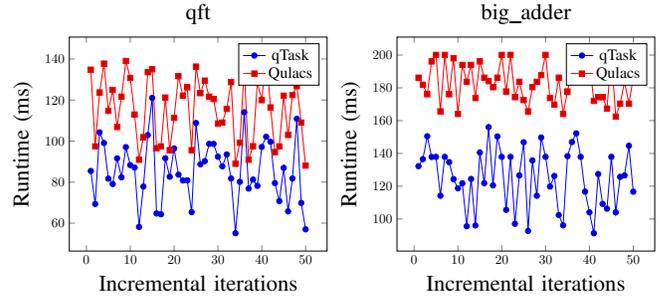
\begin{figure}[!h]
  \centering
  \pgfplotsset{
    title style={font=\LARGE},
    label style={font=\LARGE},
    legend style={font=\large},
  }
  \begin{tikzpicture}[scale=0.50]
    \begin{axis}[
      title=qft,
      ylabel=Runtime (ms),
      xlabel=Incremental iterations,
      legend pos=north east,
    ]
    \addplot+ table[x=iter,y=qtask,col sep=space]{Fig/runtime_incremental_mixes_qft.txt};
    \addplot+ table[x=iter,y=qulacs,col sep=space]{Fig/runtime_incremental_mixes_qft.txt};
    \legend{qTask, Qulacs}
    \end{axis}
  \end{tikzpicture}
  \begin{tikzpicture}[scale=0.50]
    \begin{axis}[
      title=big\_adder,
      ylabel=Runtime (ms),
      xlabel=Incremental iterations,
      legend pos=north east,
    ]
    \addplot+ table[x=iter,y=qtask,col sep=space]{Fig/runtime_incremental_mixes_big_adder.txt};
    \addplot+ table[x=iter,y=qulacs,col sep=space]{Fig/runtime_incremental_mixes_big_adder.txt};
    \legend{qTask, Qulacs}
    \end{axis}
  \end{tikzpicture}
  \caption{Performance of incremental simulation for mixing random gate removals and insertions 
  based on two quantum circuits, qft and big\_adder.}
  \label{fig::runtime_incremental_mixes}
\end{figure}

Finally, we demonstrate the performance of incremental simulation performance 
by randomly mixing gate insertions and gate removals at each incremental iteration.
Figure \ref{fig::runtime_incremental_mixes}
plots the runtime over 50 incremental iterations.
Since the circuit size varies at each iteration,
both Qulacs and qTask fluctuate.
However, we observe qTask is faster than Qulacs in nearly all points,
as a result of incremental simulation.
The runtime difference is larger at big\_adder,
which is primarily composed of non-superposition gates 
(CNOT, CX) to to perform quantum arithmetics.
In this case, qTask can quickly update certain, affected amplitudes
by circuit modifiers, rather than the entire state.

\subsection{Multi-threading Performance}

Figure \ref{fig::runtime_cores_full} compares the runtime
between Qulacs and qTask for completing full simulation using different numbers of cores.
Both qTask and Qulacs saturate at about 10 cores.
Since our task partitioning strategy enables both inter- and intra-gate operation parallelisms,
multi-threaded qTask is always faster than Qulacs.
Also, by modeling partitioned simulation tasks into a task graph,
qTask can execute the whole-graph with dynamic load balancing (via Taskflow~\cite{taskflow})
in no need of synchronizing work between levels as Qulacs.
Similar result is observed for incremental simulation in Figure \ref{fig::runtime_cores_inc},
where we collect 50 incremental iterations of random gate insertions and removals.
For qTask, the advantage of multi-threading is less significant than full simulation
because incremental simulation takes much less computation.
The scalability saturates at about 10 cores
because most task graphs give 10—-30 parallel tasks with a partition size of 256.
Smaller partition size gives more task parallelism, 
but the resulting scheduling overhead can outweigh the advantage.

\begin{figure}[!h]
  \centering
  \pgfplotsset{
    title style={font=\LARGE},
    label style={font=\LARGE},
    legend style={font=\large},
  }
  \begin{tikzpicture}[scale=0.50]
    \begin{axis}[
      title=qft (full),
      ylabel=Runtime (ms),
      xlabel=Number of cores,
      legend pos=north east,
    ]
    \addplot+ table[x=core,y=qtask-full,col sep=space]{Fig/runtime_cores_qft.txt};
    \addplot+ table[x=core,y=qulacs-full,col sep=space]{Fig/runtime_cores_qft.txt};
    \legend{qTask, Qulacs}
    \end{axis}
  \end{tikzpicture}
  \begin{tikzpicture}[scale=0.50]
    \begin{axis}[
      title=big\_adder (ful),
      ylabel=Runtime (ms),
      xlabel=Number of cores,
      legend pos=north east,
    ]
    \addplot+ table[x=core,y=qtask-full,col sep=space]{Fig/runtime_cores_big_adder.txt};
    \addplot+ table[x=core,y=qulacs-full,col sep=space]{Fig/runtime_cores_big_adder.txt};
    \legend{qTask, Qulacs}
    \end{axis}
  \end{tikzpicture}
  \caption{Runtime scalability of full simulation with increasing numbers of CPU cores
  for qft and big\_adder.}
  \label{fig::runtime_cores_full}
\end{figure}
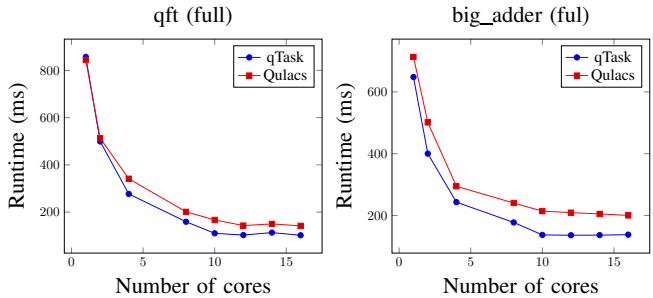
\begin{figure}[!h]
  \centering
  \pgfplotsset{
    title style={font=\LARGE},
    label style={font=\LARGE},
    legend style={font=\large},
  }
  \begin{tikzpicture}[scale=0.50]
    \begin{axis}[
      title=qft (incremental),
      ylabel=Runtime (s),
      xlabel=Number of cores,
      legend pos=north east,
      y filter/.code={\pgfmathparse{#1/1000}\pgfmathresult}
    ]
    \addplot+ table[x=core,y=qtask,col sep=space]{Fig/runtime_cores_qft.txt};
    \addplot+ table[x=core,y=qulacs,col sep=space]{Fig/runtime_cores_qft.txt};
    \legend{qTask, Qulacs}
    \end{axis}
  \end{tikzpicture}
  \begin{tikzpicture}[scale=0.50]
    \begin{axis}[
      title=big\_adder (incremental),
      ylabel=Runtime (s),
      xlabel=Number of cores,
      legend pos=north east,
      y filter/.code={\pgfmathparse{#1/1000}\pgfmathresult}
    ]
    \addplot+ table[x=core,y=qtask,col sep=space]{Fig/runtime_cores_big_adder.txt};
    \addplot+ table[x=core,y=qulacs,col sep=space]{Fig/runtime_cores_big_adder.txt};
    \legend{qTask, Qulacs}
    \end{axis}
  \end{tikzpicture}
  \caption{Runtime scalability of incremental simulation with increasing numbers of CPU cores
  for qft and big\_adder.}
  \label{fig::runtime_cores_inc}
\end{figure}
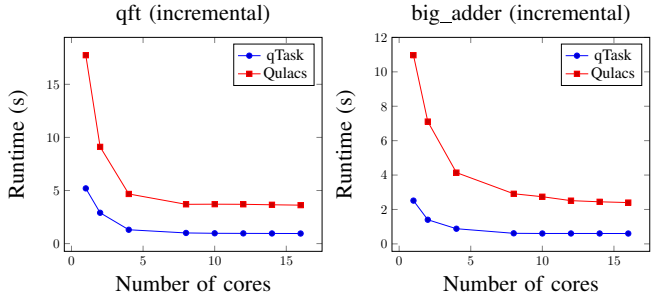

\subsection{Impact of Block Size}

\begin{figure}[!h]
  \centering
  \pgfplotsset{
    title style={font=\LARGE},
    label style={font=\LARGE},
    legend style={font=\large},
  }
  \begin{tikzpicture}[scale=0.50]
    \begin{axis}[
      title=qft (full),
      ylabel=Runtime (ms),
      xlabel=Block size ($\log B$),
      legend pos=north east,
    ]
    \addplot+ table[x=block,y=qtask-full,col sep=space]{Fig/partition_qft.txt};
    \legend{qTask}
    \end{axis}
  \end{tikzpicture}
  \begin{tikzpicture}[scale=0.50]
    \begin{axis}[
      title=qft (inc),
      ylabel=Runtime (s),
      xlabel=Block size ($\log B$),
      legend pos=north east,
      y filter/.code={\pgfmathparse{#1/1000}\pgfmathresult}
    ]
    \addplot+ table[x=block,y=qtask-inc,col sep=space]{Fig/partition_qft.txt};
    \legend{qTask}
    \end{axis}
  \end{tikzpicture}
  \caption{Runtime of full simulation and incremental simulation using different block sizes.}
  \label{fig::partition_impacts}
\end{figure}
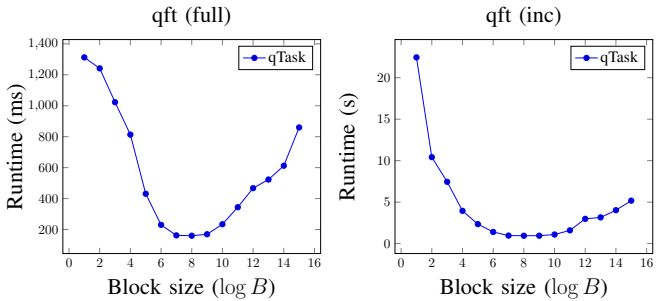

We study the impact of different block sizes on simulation performance.
In qTask, using a smaller block size results in more partitioned tasks
and thus a finer control over incrementality, and vice versa.
However, more partitions also incur higher runtime overhead,
such as re-connecting the task graph after circuit modifiers and scheduling
tasks with dynamic load balancing.
Figure \ref{fig::partition_impacts} shows the simulation runtime of
qTask for qft (15 qubits) using different block sizes.
When the block size is too small, the overhead of task partitioning and scheduling
completely outweighs the advantage of task parallelism.
When the block size is to too large, qTask does not benefit much from task parallelism,
and the result basically degenerates to using one core
(compared to Figure \ref{fig::runtime_cores_full} and Figure \ref{fig::runtime_cores_inc}).

The selection of partition size depends on the circuit structure and application environment. 
For example, if the simulator only has four cores to run, 
then a bigger partition size is better for avoiding excessive parallelism 
plus scheduling overhead. 
On the other hand, if the circuit incorporates a long chain of arithmetic operations (e.g., CNOT), 
a smaller partition size may bring more inter-gate parallelism. 
Since there is no universal optimal selection, we have decided to parameterize it for users.

\subsection{Impact of Copy-on-Write Data Optimization}

qTask partitions each state vector into a set of data blocks and 
stores each partitioned block using a C++ COW smart pointer~\cite{COW}.
A data block is automatically freed when reference count drops to zero,
which happens on the fly 
by Taskflow's scheduler.
In general, this data block management strategy can reduce the overall memory footprint by about 20--50\%.
For large-scale circuits, e.g., big\_qft and big\_ising, the saving can be
significant (up to several GBs).

\section{Conclusion}

In this paper, we have introduced qTask to efficiently support incremental quantum 
circuit simulation.
To the best knowledge of authors,
qTask is the first incremental quantum circuit simulator
in the literature.
We have presented a task-parallel decomposition strategy
to explore both inter- and intra-gate operation parallelisms
from partitioned data blocks.
Our strategy effectively scopes down
incremental update to a small set of affected partitions that can be quickly 
identified from a sequence of circuit modifiers.
We have demonstrated the promising performance of qTask on medium- and large-scale quantum circuits
from QASMBench.
Compared to two state-of-the-art simulators, Qulacs and Qiskit,
qTask is respectively 1.46$\times$ and 1.71$\times$ faster for full simulation 
and 5.77$\times$ and 9.76$\times$ faster for incremental simulation.

As part of our future work,
we are enhancing qTask to support a higher number of qubits by 
extending its state vector data structure to out-of-core memory
and distributed computing~\cite{DtCraftICCAD, DtCraft}.
Additionally, we plan to leverage the new CUDA Graph execution model~\cite{SNIG, Lin_21_01, SNIG_TPDS}
to accelerate large simulation task graphs using GPU computing.
Integrating qTask into existing quantum circuit synthesis engines~\cite{DA4QC, QubitRouting}
is also of our interest.

\section*{Acknowledgment}

We are grateful for the support of National Science Foundation (NSF) grants, 
CCF-2126672, CCF-2144523 (CAREER), OAC-2209957, and TI-2229304.
Also, the authors would like to thank reviewers for their constructive comments
on improving this paper.

\bibliographystyle{IEEEtran}

\bibliography{IEEEabrv,ms}

\end{document}